# AI-based Optimal scheduling of Renewable AC Microgrids with bidirectional LSTM-Based Wind Power Forecasting

Hossein Mohammadi[1], Shiva Jokar[2], Mojtaba Mohammadi[3], Abdollah Kavousifard[3], Morteza dabbaghjamanesh[4]

[1] Department of Electrical and Electronics Engineering, Darion Branch, Islamic Azad University, Shiraz, Iran
[2] Department of Electrical Engineering, Iranian University of Science and Technology, Iran
[3] Department of Electrical and Electronics Engineering, Shiraz University of Technology, Shiraz, Iran
[4] Electric reliability council of Texas, Austin, Texas, USA

**Abstract—** In terms of the operation of microgrids, optimal scheduling is a vital issue that must be taken into account. In this regard, this paper proposes an effective framework for optimal scheduling of renewable microgrids considering energy storage devices, wind turbines, micro turbines. Due to the nonlinearity and complexity of operation problems in microgrids, it is vital to use an accurate and robust optimization technique to efficiently solve this problem. To this end, in the proposed framework, the teacher learning-based optimization is utilized to efficiently solve the scheduling problem in the system. Moreover, a deep learning model based on bidirectional long short-term memory is proposed to address the short-term wind power forecasting problem. The feasibility and performance of the proposed framework as well as the effect of wind power forecasting on the operation efficiency are examined using IEEE 33-bus test system. Also, the Australian Wool north wind site data is utilized as a real-world dataset to evaluate the performance of the forecasting model. Results show the effective and efficient performance of the proposed framework in the optimal scheduling of microgrids.

## NOMENCLATURE

| Symbol | Description |
|---|---|
| $B_{Gi}^t / B_{si}^t$ | Cost of generation units /ESs |
| $B_{Grid}^t$ | Cost of upstream grid power |
| $b_i, b_f, b_o, b_h, \ldots$ | Bias vectors |
| $c_t$ | Cell state |
| $DR_i/UR_i$ | ramp rates |
| $f_t$ | Forget gate |
| $h(X)$ | The cost function |
| $i_t$ | Input gate at time $t$ |
| $N_T$ | Number of time intervals |
| $N_d / N_{d\text{-}DC} / N_{d\text{-}AC}$ | Number of generation units in the MG |
| $N_s$ | Number of ESs in the grid |
| $N_b$ | Number busses |
| $o_t$ | Output gate |
| $N_s / N_{s\text{-}DC}$ | Number of batteries in MG |
| $N_{Load}$ | number of loads |
| $P_{Grid,min}^t / P_{Grid,max}^t$ | Min/max power of the upstream grid |
| $P^{inj,t}_j / Q^{inj,t}_j$ | Injected active/reactive power |
| $P_{Gi}^{min} / P_{Gi}^{max}$ | Max/min i$^{th}$ generation units' output power |
| $P_{loss}^t$ | Total power loss |
| $P_{i,max}^{line,t}$ | Feeder capacity |
| $P_{Grid}^t$ | The power exchanged with the upstream grid |
| $P_{charge/discharge}$ | Permitted charge/discharge rate within a definite period $\Delta t$ |
| $P_{charge/discharge,max}$ | The maximum value of charging/discharging rate within a defined period of time $\Delta t$ |
| $P_{si}^t / P_{Gi}^t$ | Power of batteries/generation units |
| $RES^t$ | The spinning reserve |
| $S_{Gi}^{on} / S_{Gi}^{off}$ | Generation units ' Startup/shutdown cost |
| $S_{sj}^{on} / S_{sj}^{off}$ | ESs' Startup/shutdown cost |
| $u_i^t$ | ON/OFF Status of the i$^{th}$ generation units/ES |
| $V_{min}^i / V_{max}^i$ | Voltage limitations of bus |
| $V/\delta$ | Magnitude/phase of the voltage |
| $W_{ess}^t$ | battery energy storage |
| $W_{ess,max/min}$ | Capacity of ES |
| $W_{f,x}, W_{i,x}, W_{o,h}, \ldots$ | Weighting matrices |
| $X$ | Control variables |
| $Y/\Theta$ | line impedance magnitude/phase |
| $y_t$ | LSTM cells' output at time t |
| $\eta_{discharge,\ charge}$ | ES charging/discharging efficiency |
| $\sigma$ | Logistic sigmoid function |

## I. INTRODUCTION

Smart grids, which include multiple microgrids (MGs) at the distribution level, are a significant interest area in future electrical grids. As a result, to reduce costs and have a reliable operation in smart grids, MGs' energy management and optimal scheduling must be taken into account [1]. Generally, there are three classes of MGs in terms of voltage: AC MGs, DC MGs, and hybrid AC-DC MGs. Given that MGs are intermediaries between the end-users and utilities, energy management is one of the main issues. MGs operations are categorized into two different modes: grid-connected and isolated. Isolated MGs can be used to supply remote loads and maintain the system performance in the event of a disturbance in the upstream grid. However, one of the main obstacles in the isolated method is that they are not connected to the utility grid. Therefore, loads must be supplied using only the energy provided by distributed generation units and energy storages (ESs), which is the greatest challenge to manage. On the other hand, MG is connected to the utility grid through a common coupling point in the grid-connected method. This allows power to be exchanged with the utility grid in exchange for costs and benefits. Renewable energy sources (RESs) and ESs are among the technologies that have recently been considered due to environmental conditions, greenhouse gas emissions, and limited fossil fuels. Integrating RES and ESs with power grids can cause positive changes and benefits. Additionally, old distribution grids have changed due to new technologies such as ESs and RESs. That is why MGs are so important as the next generation of the power grid, and the most critical issue in the performance of MGs is their optimal energy management and optimal scheduling. Today, wind turbines have become one of the most popular technologies. As mentioned before, RESs are among future trends in smart grids that must be studied. As a RES, wind turbines are one of the fastest ways to generate electricity from renewable energy. Due to the many benefits of this technology, scientists have done much research to eliminate the challenges in this technology, including increasing the accuracy of forecasting in the system as much as possible. In this regard, wind power forecasting is one of the main topics of this study.

In recent years, energy management and optimal scheduling of MGs have received considerable researchers' attention. Authors in [2] have proposed using a probabilistic energy management scheme based on the crow search method to minimize the cost of the hybrid MGs (HMGs). That paper has focused on using unscented transform to deal with uncertain variables. In [3], an energy management scheme is developed for combined heat power (CHP) based MGs as a multi-objective optimization problem. Authors in [4], suggested the 2m point estimation algorithm for MGs to develop a stochastic management framework. Authors in [5], presented a management plan considering electric vehicles, ESs. Paper [6] proposes an energy management method for CHP-based MGs that takes demand response technology and a variety of generating units into account. In that paper, demand response is utilized for improving energy management and generative adversarial nets for renewable power forecasting. A machine learning-based energy management system for optimal scheduling of HMGs has been suggested in [7]. Reference [8] has investigated the components and basics of HMGs. in that paper, the energy management of CHP-based MGs is formulated as a multi-objective optimization problem, and different types of units, as well as demand response technology, are utilized. In Article [9], different generation units and uncertainties in the system are considered and a statistic energy policy using odorless conversion for MGs is proposed. The article [10] proposed a stochastic energy management method for AC/DC MGs. Various generating units are addressed in that work, and system uncertainties are recorded using the unscented transform. Paper [11], suggested an energy management technique for CHP-based MGs that took into account the reaction power demand and various generating units. The reference [5], suggested a management scheme for HMGs that took into account fuel cell CHP models, plug-in electric vehicles, ESs, RESs, and a strategy for feeder reconfiguration. A secure distributed cloud-fog-based optimal scheduling system for MGs was presented in reference [12]. A two-layer energy management method for MGs with high penetration of RESs and ESs considering battery degradation is proposed in [13]. Reference [14] provides an efficient energy management technique for renewable MGs considering ESs and tidal power generation units. While the above works addressed significant topics regarding the energy management of MGs, the research in this area is still in its infancy. In this regard, this paper proposes a robust energy management of MGs considering renewable energy sources.

Another vital concept in modern power grids is Renewable energy. This concept, which is essential to power grids, consists of several concepts such as photovoltaics and wind turbines. Due to the vital role of wind power in optimal scheduling and operation on power grids, wind power forecasting has attracted many researchers in recent years. Reference [15] suggested a three-layer backpropagation (BP) neural network (NN) prediction method for one-step wind farm wind speed prediction. In [16], a prediction technique based on the wavelet transform for wind power forecasting is proposed. Future wind data are forecasted using this strategy by analyzing hypothetical symptoms using a backpropagation neural network. In[6], a forecasting model based on generative adversarial networks for renewable units is proposed and the results of the proposed method are compared with several well-known methods.

Reference [17] suggested a recurrent neural network (RNN) based method to estimate wind power by learning the temporal relationship inherent in the time series. RNNs are a subclass of artificial neural networks. Their unique internal state, which contains memory for old information, enables RNNs to create a directed graph along a series and to learn the temporal dynamic response of a time series. Reference [18] recommended a forecasting model using long short-term memory (LSTM) to address gradient vanishing problem and explosion. As one of the advanced variations of recurrent neural networks, LSTM can more successfully learn the information stored in time series data. In order to improve load forecasting issue in MGs, a novel method based on bidirectional LSTM (BLSTM) for wind power forecasting is proposed in this paper.

While the above works addressed important topics that clarify the importance of energy management and wind power forecasting in MGs, the research in this area is still in its infancy. Earlier research has shown a continuous and stable relationship between MGs and power grids. However, data on the operation of MGs are limited and prior studies have suffered from methodological flaws in their case selection. By calculating the amount of electricity consumed by the grid at various times of the day and considering the cost of electricity generation by each of the fossil fuel and renewable energy power plants in the MGs, this study determines the optimal model for generating electricity at various times in the grid from 24 hours ago. This method is designed to minimize the MGs' operation cost. This study has two key goals. First, to provide an optimal energy management strategy, and second, to address wind power forecasting problems in MGs using BLSTM. In this research, wind turbines, fuel cell units, micro turbines, and ESs are considered. This paper offers a powerful optimization method called teacher learning-based optimization (TLBO) for solving the system's scheduling problem. The feasibility of the proposed energy management strategy is examined using an IEEE test system and the performance of the wind forecasting technique is investigated using real-world data.

In section II, the optimal scheduling and operation of the MG are formulated as an optimization problem, and the details of the TLBO are presented. Section III is focused on the proposed deep-learning-based forecasting model. The simulation results and conclusion are provided in sections IV and V.

## II. MG OPERATION AND OPTIMAL SCHEDULING

MGs are designed to facilitate electricity power supply at the distribution level. Fig. 1 illustrates the schematic construction of typical MGs. In smart MGs, physical layer parameters (e.g. voltage, active power, and reactive power, etc.) are measured and sent to the MG central control at regular time intervals through advanced metering infrastructures (AMI). Each node in these systems is equipped with a smart metering device that communicates with central control through communication channels. for one day ahead of the operation, MG central control schedules the system according to the system's characteristics and the data of energy market price, renewable energy sources' forecasted output power, forecasted load demand. In the following, the operation of MG is formulated as a constraint optimization problem and then the TLBO is described in detail.

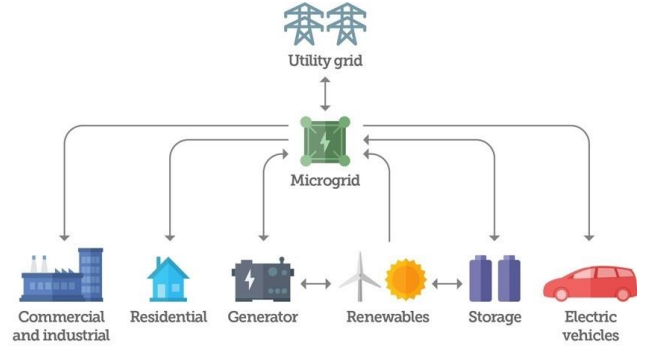

Fig. 1 schematic construction of MG's structure

### A. MG Scheduling Problem Formulation

In this section, the operation of renewable MGs is modeled as a single objective optimization problem. The main cost objective function incorporates the cost of generation units, ESs, and power sold/ purchased to/ from the utility grid as follows [19]:

$$Min \ h(X) = \sum_{t=1}^{N_T}(\sum_{i=1}^{N_g}[u_i^t P_{Gi}^t B_{Gi}^t + S_{Gi}^{on} \max\{0, u_i^t - u_i^{t-1}\} +$$

$$+S_{Gi}^{off} \max\{0, u_i^{t-1} - u_i^t\}] + \sum_{j=1}^{N_S}[u_j^t P_{sj}^t B_{sj}^t + S_{sj}^{on} \max\{0, u_j^t - u_j^{t-1}\} + \quad (1)$$

$$+S_{sj}^{off} \max\{0, u_j^{t-1} - u_j^t\}] + P_{Grid}^t B_{Grid}^t)$$

The variable X in (1) indicates the optimum operation point of grid components and their binary OFF/ON status as follow:

$$X = [P_g, U_g]_{1\times(2\times n\times N_T)} \ , \ n = N_d + N_s + 1 \ ; \ \forall t \in N_T$$
$$P_g^t = [P_G^t, P_s^t, P_{Grid}^t] \ ; \ P_G^t = [P_{G1}^t, P_{G2}^t,...,P_{GN_g}^t]$$
$$U_g^t = [u_1^t, u_2^t,...,u_{N_d}^t] \ ; \ P_s^t = [P_{s1}^t, P_{s2}^t,...,P_{sN_s}^t] \quad (2)$$
$$P_{Grid}^t = [P_{Grid}^t] \ , \ u_k^t \in \{0,1\}$$

The primary optimization challenge is solved by taking into account various practical limitations as follows. The balance between consumption and generation in the system is maintained through the following load flow equations:

$$P_j^{inj,t} = \sum_{n=1}^{N_b} V_j^t V_n^t Y_{jn} \cos(\theta_{jn} + \delta_j - \delta_n) \quad (3)$$

$$Q_j^{inj,t} = \sum_{n=1}^{N_b} V_j^t V_n^t Y_{jn} \sin(\theta_{jn} + \delta_j - \delta_n) \quad (4)$$

The restriction on the generation units, utility grid, and power of converters are presented in the rest.

$$P_{Gi,\min}^t \leq P_{Gi}^t \leq P_{Gi,\max}^t$$
$$P_{conv,\min}^t \leq P_{conv}^t \leq P_{conv,\max}^t \quad (5)$$
$$P_{Grid,\min}^t \leq P_{Grid}^t \leq P_{Grid,\max}^t$$

-feeder capacity:
$$|P_i^{Line,t}| \leq P_{i,\max}^{Line} \quad (6)$$

-spinning reserve:
$$\sum_{i=1}^{N_d} u_i^t P_{Gi,\max}^t + P_{Grid,\max}^t \geq \sum_{k=1}^{N_{Load}} P_{Load,k}^t + P_{loss}^t + \text{Res}^t \quad (7)$$

-bus voltage:
$$V_m^{\min} \leq V_m^t \leq V_m^{\max} \quad (8)$$

- generation units' ramp rate:
$$|P_{Gi}^t - P_{Gi}^{t-1}| < UR_i, DR_i \quad (9)$$

-energy storage constraints:
$$W_{ess}^t = W_{ess}^{t-1} + \eta_{charge} P_{charge} \Delta t - \frac{1}{\eta_{discharge}} P_{discharge} \Delta t \quad (10)$$

$$\begin{cases} W_{ess,\min} \leq W_{ess}^t \leq W_{ess,\max} \\ P_{charge,t} \leq P_{charge,\max} \\ P_{discharge,t} \leq P_{discharge,\max} \end{cases} \quad (11)$$

B. Teacher-learning based optimization algorithm

As mentioned before, the TLBO is utilized to solve the scheduling problem of MGs and minimize the operation cost. Due to the high ability of evolutionary optimization techniques in solving constraint/non-constraint single/multi-objective optimization problems, these algorithms have attracted a lot of attention in recent years. For instance, authors in [2] used the crow search optimization to solve the optimal scheduling problem of HMG respectively. In this paper, the TLBO algorithm is employed to solve the optimal scheduling problem of MGs.

The TLBO algorithm is a population-based algorithm that models the learning process of students in the class [20]. The algorithm is consisting of two phases: the teacher phase and the student phase. The main idea behind the teacher phase is that during the learning process teacher attempts to move the mean grade of students toward the best student. It is worth noting that in each iteration, the teacher is considered the best current solution. The mathematical model of the teacher phase is described as follows:

$$\Delta X_{iT} = r_i (T_i - F_i M_i) \quad (13)$$
$$X_i^{new} = X_i^{old} + \Delta X_{iT} \quad (14)$$

Where $T_i$, $M_i$, $r_i$, and $F_i$ present the teacher, mean grade of students, a random number in the range of [0,1], and a random integer between *1* and *2* in the $i^{th}$ iteration, respectively. The second phase (i.e. student phase) simulates the learning process among students in which students increase their knowledge by discussing with each other. For two randomly chosen students, the student phase is expressed as follows:

$$\begin{cases} X_i^{new} = X_i^{old} + r_i(X_i - X_j) & \text{if } F(X_i) < F(X_j) \\ X_i^{new} = X_i^{old} + r_i(X_j - X_i) & \text{if } F(X_i) > F(X_j) \end{cases} \quad i \neq j \quad (15)$$

Note that, in (13) the *F(X)* presents the cost objective function of students. The advantages of TLBO over other well-known optimization methods (e.g. PSO, CSA) are its simplicity, two-step movement in search space, and no dependence on regulatory parameters.

III. WIND POWER FORECASTING MODEL BASED ON BLSTM

In this section, the details of the proposed wind power forecasting technique are provided. The proposed method is designed such that it can forecast the future wind power only using previously measured values. In other words, the proposed model doesn't require any additional information such as wind speed, temperature, etc. The first stage of the proposed model is the dataset generation stage. In this stage, first, the historical data is processed to generate the proper dataset for training the BLSTM model.

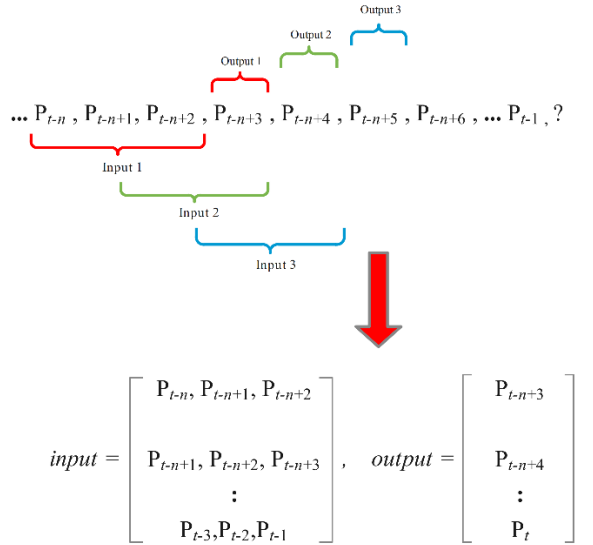

Fig. 2 Dataset generation stage

Fig. 2 shows the dataset generation stage. In this stage, two windows of size *m* and *n* are utilized to generate the input and the corresponding outputs from the sequence of historical data. For instance, if m is 4 and n is 1, the input related to each data sample is its four previous samples. In Fig. 2, m = 3 and n = 1 although, the window size can vary for different situations. After the dataset is generated in the first stage, a BLSTM is trained. In the following, the details of the BLSTM are presented.

LSTMs were first proposed as a solution to the gradient vanishing issue in regular RNN [21]. The LSTM records and stores information by using the cell state and three cell gates. The input and forget gates indicate the data to be added to and removed from the cell state, respectively. Additionally, the output gate determines which section of the cell state should be output. Fig. 3 depicts the construction of the LSTM cell. Take note that the LSTM network's training procedure is based on backpropagation [22]. The LSTM's transition equations are as follows:

$$h_t = H(W_{x,h}x_t + W_{h,h}h_{t-1} + b_h) \quad (16)$$

$$y_t = W_{h,y}h_t + b_y \quad (12)$$

The following equations implement the function H:

$$i_t = \partial(W_{i,x}x_t + W_{i,h}h_{t-1} + b_i) \quad (17)$$

$$f_t = \partial(W_{f,x}x_t + W_{f,h}h_{t-1} + b_f) \quad (18)$$

$$\overline{c_t} = \tanh(W_{\overline{c},x}x_t + W_{\overline{c},x}h_{t-1} + b_{\overline{c}}) \quad (19)$$

$$o_t = \partial(W_{o,x}x_t + W_{o,h}h_{t-1} + b_o) \quad (20)$$

$$C_t = f_t \cdot c_{t-1} + i_t \cdot \overline{c_t} \quad (21)$$

$$h_t = o_t \cdot \tanh(c_t) \quad (22)$$

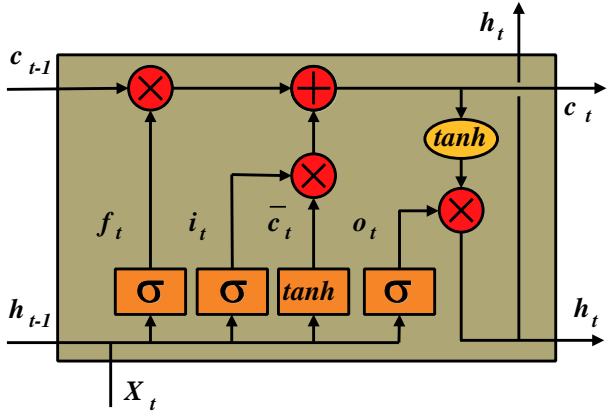

Fig. 3 LSTM cell structure

The notion of memory cells with regulating gates enables LSTMs to significantly outperform RNNs in overcoming long-term reliance and vanishing gradients [21]. Through backpropagation, the learning process calculates the weights that determine whether the cells store or delete data.

Another variant on RNNs is presented in [23], in which a single RNN layer is composed of two RNN blocks processing temporal input in opposing directions concurrently. At each time occurrence, the ultimate output is a sum of the results of each RNN block. As seen in Fig. 3, the bidirectional structure could be applied to LSTM to create a BLSTM. In the following, the mathematical relation between forwarding and backward processes and the network's output is described:

$$\vec{h}_t = H(W_{x,\vec{h}}x_t + W_{\vec{h},\vec{h}}\vec{h}_{t-1} + b_{\vec{h}}) \quad (23)$$

$$\overleftarrow{h}_t = H(W_{x,\overleftarrow{h}}x_t + W_{\overleftarrow{h},\overleftarrow{h}}\overleftarrow{h}_{t-1} + b_{\overleftarrow{h}}) \quad (24)$$

$$y_t = W_{\vec{h},y}\vec{h}_t + W_{\overleftarrow{h},y}\overleftarrow{h}_t + b_y \quad (25)$$

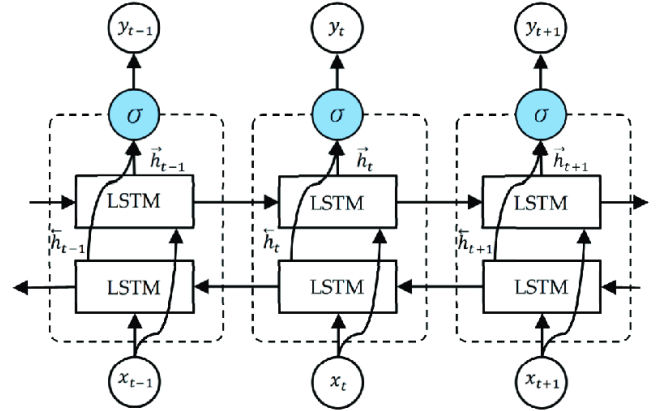

Fig. 4 BLSTM structure

## IV. SIMULATION RESULTS

In this section, the performance and feasibility of the proposed strategies are examined on practical test cases that are constructed based on the IEEE 33-bus test system. The test system includes a wind turbine, two micro turbines, and one ES. The complete details of the characteristics of the components can be found in Table I. The test system operates based on the 12 kV voltage level and the MG is connected to the upstream grid through the point of common coupling at bus 1 and also MG and the upstream grid can exchange power based on cost and benefits. In order to respect the idea of green energy, the wind turbine unit is considered non-dispatchable meaning that all of its generations are consumed without considering the cost-efficiency. In this study, the complete model for transmission lines (impedance) is considered. The hourly electricity market price and load factor are provided in figures 5 and 6 [24].

Moreover, the historical data of the Woolnorth wind site [25], which is located at Woolnorth on Cape Grim at the northwest tip of Tasmania, is utilized to evaluate the performance of the proposed wind power forecasting model. It is worth noting that Woolnorth wind site is one of the most difficult situations for wind power forecasting in Australia due to its location on the edge of a cliff exposed to the Southern Ocean. The data set include the one-hour instantaneous readings in MW from 30 December 2009 to 1 January 2010. In the following, first, the proposed wind power forecasting model is examined, and then its results are utilized to evaluate the performance of the proposed optimal scheduling strategy.

Table I characteristics of the components of the system

| Type | Min power (kW) | Max power (kW) | Bid ($/kWh) | Startup/shutdown cost ($ct) | Ramp rate | Location (Bus number) |
|---|---|---|---|---|---|---|
| Micro-turbine 1 | 100 | 1300 | 0.645 | 75 | 220 | 12 |
| Micro-turbine 2 | 90 | 1100 | 0.675 | 70 | 180 | 25 |
| Wind-turbine 1 | 0 | 4000 | 1.073 | 0 | - | 30 |
| Energy storage | -250 | 250 | 0.318 | 0 | - | 18 |

### A. Wind power forecasting results

As mentioned before, the Woolnorth wind site hourly historical data is utilized to evaluate the performance of the proposed forecasting model. For this purpose, the data is divided to test and train datasets where the split ratio for the train is 0.8. The dataset generation stage is performed on data considering m=48 and n=24 meaning that the proposed model can forecast the next 24 hours of data simultaneously. The proposed model is implemented in python 3.8 using Tensorflow.Keras library. The optimizer of the model is ADAM, the number of epochs is 250, and the loss function is mean square error. The network architecture includes two 128-cell layers, one dropout layer with a ratio of 0.3, and one fully connected layer. The activation function of hidden layers is Relu and for the last layer is linear. It is worth noting that the dropout layer is for overfitting prevention and also the 2.4 last months of the year are used for testing purposes. Figure 5 shows the regression diagram and compares the actual output power with the predicted values for 24 hours related to 22/11/2009. Note that for better evaluation, the proposed methodology is compared with two well-known methods called artificial neural network (ANN) and (LSTM).

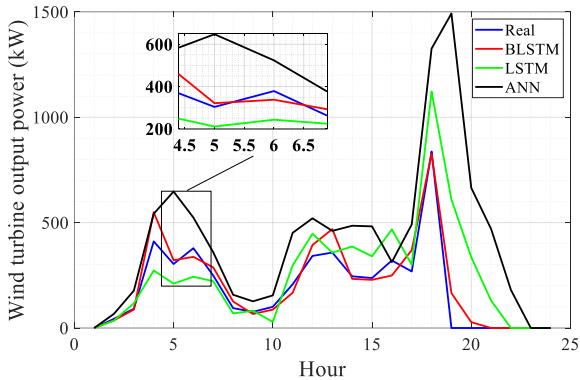

Fig. 5 Regression diagram

As can be seen from Figure 5, the proposed BLSTM-based model has a relatively better performance in comparison with ANN and LSTM. To be more specific, as it is highlighted in the figure, the real output power of the wind turbine at the 5$^{th}$ hour is 303.76 and the forecasted value of BLSTM is 321.41 where the output of LSTM and ANN are 210.84 and 647.68 respectively. Also, it can be seen that ANN has very poor performance compared to other models. Additionally, the regression metrics of the models are presented in Table II. the regression metrics in this study are mean absolute percentage error (MAPE), root mean square error (RMSE) and mean absolute error (MAE) [6]. The better performance and priority of the BLSTM model over ANN and LSTM can be easily observed in Table II such that the BLSTM is the leading model in all metrics. Also, it can be seen that, although the BLSTM has the best performance the LSTM is acceptable as well.

Table II Prediction metrics

| | MAPE | MAE | RMSE |
|---|---|---|---|
| ANN | 17.868 | 209.36 | 362.09 |
| LSTM | 9.845006 | 115.91 | 198.88 |
| BLSTM | 8.57619 | 101.07 | 173.5 |

### B. MG Optimal scheduling

This section is focused on the optimal scheduling of the MG for 24 hours. In order to highlight the effect of the wind power forecasting model on the accuracy and efficiency of the optimal scheduling, four scenarios are considered as below:

- Scenario 1: optimal scheduling is performed using real wind data.
- Scenario 2: optimal scheduling is performed using BLSTM results.
- Scenario 3: optimal scheduling is performed using LSTM results.
- Scenario 4: optimal scheduling is performed using ANN results.

Table III Power loss, maximum voltage deviation, and total cost of the grid for different scenarios

| Case study | Power loss (kW) | Maximum voltage deviation (pu) | Total cost ($) |
|---|---|---|---|
| Scenario 1 | 19837 | 0.0042 | 55440 |
| Scenario 2 | 20945 | 0.0053 | 55316 |
| Scenario 3 | 19777 | 0.0029 | 56758 |
| Scenario 4 | 19204 | 0.0023 | 58023 |

The power loss, maximum voltage deviation, and total cost of the grid for all scenarios are presented in Table III. As can be seen from that figure, the grid's total cost for scenario 2 has the least difference with scenario 1 (124$) where this value for scenarios 3 and 4 are 1318$ and 2583$. This shows the high impact of the wind power forecasting model on the optimal scheduling and energy management. It is worth noting that the maximum voltage deviation in Table IV is the highest value among all buses during the operation day. As can be seen the maximum voltage deviation for all scenarios are less than the corresponding threshold. Since the scenario 2 had the best results, from now on only the results of scenario 2 are analyzed.

Table IV shows the MG optimal scheduling for 24 hours using BLSTM results considering one hour intervals. Also, the amount of power exchanged with upstream grid and the hourly operation cost are presented in Fig. 8 and Fig. 9 respectively. According to Table I and Fig. 7, the electricity price varies at different hours of the day and also the generation cost of different units are not the same. Therefore, logically it is more beneficial to purchase the power from the upstream grid at the early hours of the day and turn of expensive units. As can be seen from Table III, dispatchable units (i.e. micro turbines 1 and 2) are turned off at the early hours of the day. On the other hand, in the middle of the day, which the electricity price is very high, all generation units operate at their maximum capacity to keep the operation cost as low as possible.

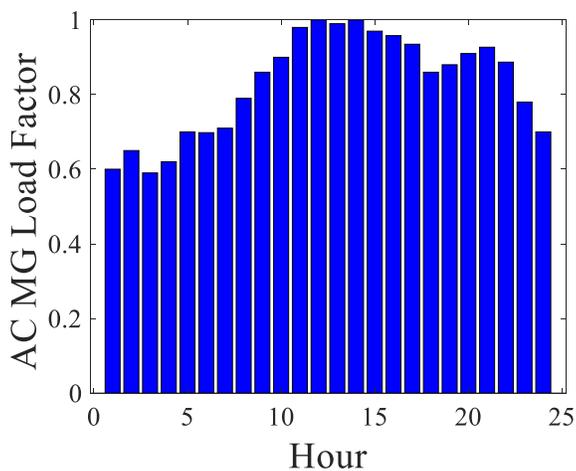

Fig. 6 hourly load factor

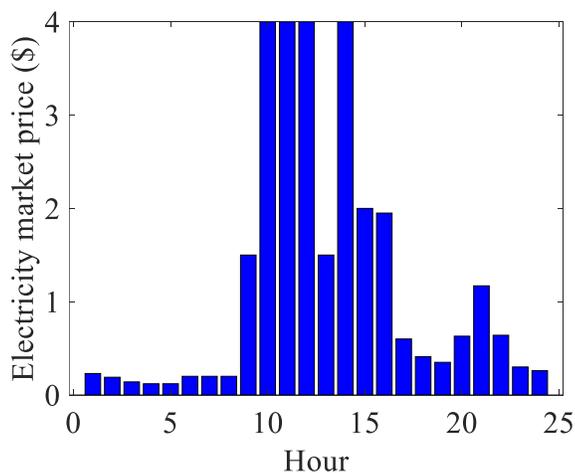

Fig 7 hourly electricity market price

Another beneficial policy for the grid would be storing energy in ES at low electricity price hours and injecting that energy to the grid in the middle of the day when the energy price is high. As can be seen from Table IV and Fig. 7, the ES is charged at the low price hours (i.e. 00:00 AM to 9:00 AM and 5:00 PM to 7:00 PM) and discharged at high price hours. This policy could decrease the total operation cost of the system significantly. Therefore, the presence of ESs in the grid can highly improve the system's efficiency.

Table IV MG optimal scheduling for 24 hours

| Hour | Energy storage | Micro turbine 1 | Micro turbine 2 | Wind Turbine |
|---|---|---|---|---|
| 1 | -250 | 0 | 0 | 0.0 |
| 2 | -250 | 0 | 0 | 36.5 |
| 3 | -250 | 0 | 0 | 87.4 |
| 4 | -250 | 0 | 0 | 549.5 |
| 5 | -250 | 0 | 180 | 321.4 |
| 6 | -250 | 220 | 360 | 337.7 |
| 7 | -250 | 440 | 540 | 287.4 |
| 8 | -250 | 660 | 720 | 125.8 |
| 9 | 250 | 880 | 900 | 66.5 |
| 10 | 250 | 1100 | 1080 | 87.3 |
| 11 | 250 | 1300 | 1100 | 167.2 |
| 12 | 250 | 1300 | 1100 | 394.0 |
| 13 | 250 | 1300 | 1100 | 469.2 |
| 14 | 250 | 1300 | 1100 | 233.2 |
| 15 | 250 | 1300 | 1100 | 228.7 |
| 16 | 250 | 1300 | 1100 | 249.3 |
| 17 | -250 | 1080 | 1100 | 369.5 |
| 18 | 250 | 860 | 920 | 827.0 |
| 19 | -250 | 640 | 740 | 165.8 |
| 20 | 0 | 440 | 920 | 27.6 |
| 21 | 250 | 660 | 740 | 0.0 |
| 22 | -250 | 440 | 560 | 0.0 |
| 23 | -250 | 220 | 380 | 0.0 |
| 24 | -250 | 0 | 200 | 0.0 |

Figures 8 and 9 show the amount of power exchanged with the upstream grid and the hourly operation cost of the grid during the operation day. According to Fig. 5 and Table 7, the amount of power exchanged with the upstream grid has a reverse relation with the energy market price, meaning that in low price hours the priority is given to the upstream grid rather than generation units. On the other hand, in the middle of the day, which electricity price is high, most of the grid's demand is supplied through generation units.

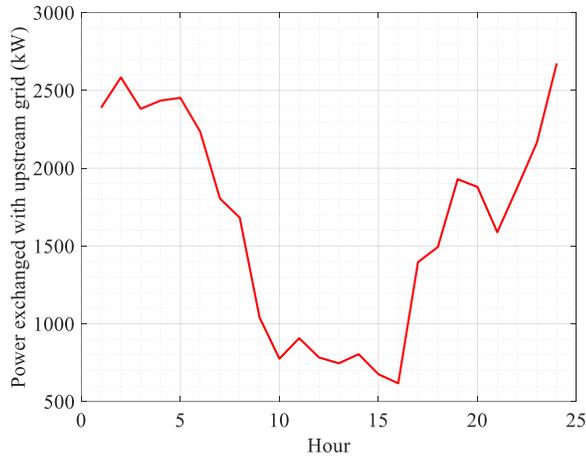

Fig. 8 the amount of power exchanged with the upstream grid

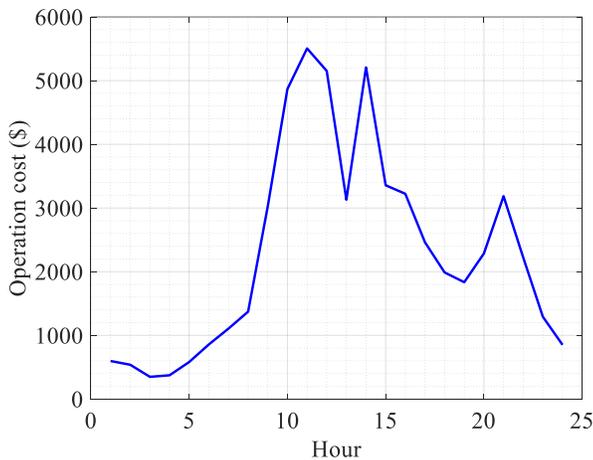

Fig. 9 Hourly operation cost of the grid

## V. CONCLUSION

This paper was focused on wind power forecasting and the optimal scheduling of renewable MGs. For robust and efficient scheduling of MG, an optimal framework based on TLBO was proposed. The first part of the proposed methodology, which is for optimal scheduling, takes the practical limitation of the system such as ramp rate, etc. to account and presents an optimal framework for efficient operation of MGs. The performance of the proposed methodology was tested on the IEE 33-bus test system including threem generation units. The results showed that the proposed methodology can efficiently minimize the operation cost through managing the generation units and batteries. Also, the impact of wind power forecasting accuracy on the efficiency of the grid was examined. The second part of the proposed framework was focused on wind power forecasting using BLSTM. The performance of the proposed model was evaluated using a real-world test case and The results showed the acceptable performance of the proposed model and its priority of over two well-known forecasting models.